\documentclass[12pt,preprint]{aastex}
\usepackage{graphicx} 
\usepackage{color}

\shortauthors{Raso et al.}

\def\ltsima{$\; \buildrel < \over \sim \;$}
\def\gtsima{$\; \buildrel > \over \sim \;$}
\def\lsim{\lower.5ex\hbox{\ltsima}}
\def\gsim{\lower.5ex\hbox{\gtsima}}

\begin{document}

\title{The ``UV-route" to search for Blue Straggler Stars in Globular
  Clusters: first results from the HST UV Legacy Survey}

\author{S. Raso$^{1,2}$, F.R. Ferraro$^1$, E. Dalessandro$^2$,
  B. Lanzoni$^1$, D. Nardiello$^3$, A. Bellini$^4$,
  E. Vesperini$^5$ }

\affil{
\textsuperscript{1} Dept. of Physics and Astronomy, University of
Bologna, Viale Berti Pichat, 6/2, Bologna, Italy\\
\textsuperscript{2} INAF Osservatorio Astronomico di Bologna, Via
Ranzani 1, Bologna, Italy\\
\textsuperscript{3} Dept. of Physics and Astronomy Galileo Galilei, University of 
 Padova, Vicolo dell'Osservatorio 3, I-35122 Padova, Italy)\\
\textsuperscript{4} Space Telescope Science Institute, 3700 San Martin Drive, Baltimore, MD
21218, USA.\\
\textsuperscript{5} Dept. of Astronomy, Indiana University,
Bloomington, IN, 47401, USA.}

\date{20 March 2017}

\begin{abstract}
We used data from the HST UV Legacy Survey of Galactic Globular
Clusters to select the Blue Straggler Star (BSS) population in four
intermediate/high density systems (namely NGC 2808, NGC 6388, NGC 6541
and NGC 7078) through a ``UV-guided search''.  This procedure consists
in using the F275W images in each cluster to construct the master
list of detected sources, and then force it to the images acquired in
the other filters.  Such an approach optimizes the detection of
relatively hot stars and allows the detection of complete sample of
BSSs even in the central region of high-density clusters, because the
light from the bright cool giants, which dominates the optical
emission in old stellar systems, is sensibly reduced at UV
wavelengths.  Our UV-guided selections of BSSs have been compared to
the samples obtained in previous, optical-driven surveys, clearly
demonstrating the efficiency of the UV approach.  In each cluster we
also measured the parameter $A^+$, defined as the area enclosed
between the cumulative radial distribution of BSSs and that of a
reference population, which traces the level of BSS central
segregation and the level of dynamical evolution suffered by the
system. The values measured for the four clusters studied in this paper
nicely fall along the dynamical sequence recently presented for a sample of 25
clusters.
\end{abstract}


\section{Introduction}
\label{intro}
In the color-magnitude diagram (CMD) of a stellar population, the
so-called Blue Straggler Stars (BSSs) appear as a sparse group of
stars with relatively low luminousity (a few $L_\odot$) and high
effective temperature ($6500-9000$K), lying along the extrapolation of
the main sequence (MS), at brighter magnitudes and bluer colours with
respect to the MS-turnoff (MS-TO) point (e.g., \citealp{sandage53,
  ferraro+92, ferraro+93, ferraro+95, ferraro+04}, \citealp{piotto+04,
  leigh07, moretti08}, \citealp{dalessandro+08, beccari+11,
  beccari+12, simunovic16}).  Their anomalous position in the CMD
indicates that they are more massive than the other visible stars in
the cluster, as it is also confirmed by direct mass estimates
\citep{shara97, gilliland98, ferraro+06b, fiorentino+14, baldwin+16}. Hence, BSSs
must have formed through ``exotic'' channels, as mass-transfer in
binary systems \citep{mccrea1964} and stellar mergers due to direct
collisions \citep{hills+1976}.  The full characterization of the BSS
formation mechanisms and successive evolution \citep[e.g.,][]{sills09,
  ferraro+16} and their relative importance in different enviroments
have not been completely quantified yet \citep[e.g.,][]{davies04,
  sollima08, chen09, knigge09, ferraro+03, leigh13, chatterjee13_bss,
  sills13}, but a few pieces of evidence suggest that both channels
can be active within the same cluster \citep{ferraro+09,
  dalessandro+13_n362, simunovic14,xin15}.

Because of their anomalous position in the CMD, BSSs are, at least in
principle, clearly distinguishable from the other cluster stars.
However, the construction of complete samples of BSSs can be critical
in old star clusters, since the optical emission in these systems is
primarily dominated by a large population of much brighter (several
$10^2 \times L_\odot$) and cool ($3500-5000$ K) giants.  Clearly, this
task is particularly difficult (even with the {\it Hubble Space Telescope} - HST) 
in very crowded enviroments, as those observed in the central regions of
high-density (several $10^{5-6}$ stars per cubic parsec) globular
clusters (GCs).  However, because of their hot surface temperatures,
BSSs appear among the brightest objects at UV wavelenghts, where red
giants are, instead, particularly faint. Hence the combination of high
angular resolution and UV imaging capabilities (as offered by 
the HST)
is the only solution to significantly reduce the level of crowding and
acquire complete samples of BSSs even in the core of highly
concentrated GCs.  Within this framework, approximately 20 years ago
we promoted the so-called {\it UV route to the study of BSSs in GCs}
(see \citealt{ferraro+97, ferraro+99, ferraro+01, ferraro+03}), an
approach that allowed us to derive complete samples of BSSs in several
Galactic GCs (GGCs), including systems of very high central density
(see \citealp{lanzoni+07a, lanzoni+07b, lanzoni+07c, dalessandro+08,
  dalessandro+09, sanna+12, sanna+14, contreras+12}).

The dataset recently acquired within the HST UV Legacy Survey of GGCs
(see Sect. \ref{data}; \citealp{piotto+15}) allows the extension of
this approach to a significantly larger number of clusters. While the
complete study of the BSS population in all the surveyed GCs will be
discussed in a forthcoming paper (Francesco R. Ferraro et al. 2017, in
preparation), here we present the first results obtained for four
intermediate/high density systems. The specific aim of this paper is
to illustrate the advantages of the {\it ``UV-guided''} search, with
respect to the {\it ``optical-driven''} selection
(e.g. \citealt{soto16}), for the identification of complete samples of
BSSs. A detailed comparison with the results obtained from currently
available online catalogs is therefore presented and discussed.  We
finally discuss the radial distributions of BSSs and their level of
segregation relative to HB or RGB stars; the BSS central segregation
is measured by means of the $A^+$ parameter recently defined in
\citet{alessandrini+16} and \citet{lanzoni+16}, to explore the use of
BSSs as tracers of the dynamical evolution of the parent cluster
\citep{ferraro+12, ferraro+15}.

\section{The photometric database and data reduction}
\label{data}
 The data used in this paper have been secured as part of the Hubble
 Space Telescope UV Legacy Survey of GGCs (GO-13297; PI: Piotto) and
 in two pilot companion projects (GO-12311 and GO-12605; see
 \citealt{piotto+15}). In all the 57 clusters deep images in
 the F275W, F336W and F438W filters have been acquired with the
 WFC3/UVIS.  While this filter combination provides the ideal set to
 generate pseudo-colors able to maximize the splitting of multiple
 stellar populations in the CMD (see e.g. \citealt{milone+13}), it also offers the
 possibility to construct  diagrams where an appropriate selection of
 hot populations (like BSSs) can be performed.  In particular, to
 optimally study BSSs we adopted a ``pure'' UV diagram: the ($m_{\rm
   F275W}, m_{\rm F275W}-m_{\rm 336W}$) CMD.

The four clusters selected for this paper (namely, NGC 2808, NGC 6388,
NGC 6541 and NGC 7078) are massive ($M_V<-8.5$, $M> 5\times 10^5
M_{\odot}$), with high central density ($\log \rho_0>4.6$ in units of
$L_\odot$ pc$^{-3}$)\footnote{Two of them (namely NGC 7078 and NGC
  6541) are catalogued as core collapsed clusters.} and moderately
extended core radius ($7\arcsec<r_c<15\arcsec$), and they span a large
range in metallicity ($-2.4<$[Fe/H]$<-0.55$; all parameters are from
\citealp{harris96}, 2010 edition).  Hence, the four selected clusters
represent the ideal systems to test the advantages of BSS searches in
the UV domain over those carried out using the optical band.

The WFC3/UVIS camera consists of two twin chips, each of
$4096\times2051$ pixels and with a pixel scale of $0.0395\arcsec$,
resulting in a total field of view of $\sim 162\arcsec \times
162\arcsec$.  For each cluster several images were obtained in the
F275W and F336W bands (see Table \ref{tab:exp} for details).
To allow a better subtraction of CCD defects, artifacts 
and false detections, different pointings dithered by several pixels,
and in some cases also rotated by $\sim 90 \deg$ (NGC 6541) or $\sim
70 \deg$ (NGC 6388), have been acquired in each band.
 
 For the photometric analysis we used the set of images processed,
 flat-fielded, bias subtracted and corrected for charge transfer
 efficiency (CTE) by standard HST pipelines
 (\texttt{$\textunderscore$flc} images). Pixel-area effects have been
 accounted for by applying the most updated pixel-area-maps
 (\texttt{PAM} images) to each image by means of IRAF
 tasks.\footnote{IRAF is distributed by the National Optical Astronomy
   Observatory, which is operated by the Association of Universities
   for Research in Astronomy, Inc., under cooperative agreement with
   the National Science Foundation.}  The analysis has been performed
 independently on each chip by using \texttt{DAOPHOT IV}
 \citep{stetson87}. For each image we selected several ($\sim 200$)
 bright, unsaturated and relatively isolated stars to model the
 point-spread function (PSF), for which we used a spatially variable
 Moffat function. A first star list has been obtained for each frame
 by independently fitting all the star-like sources at $3\sigma$ above
 the local background.  To take advantage of the reduced crowding
 conditions at UV wavelegths, we created a {\it master list} composed
 of stars detected in a sub-sample of at least half of the total
 number of F275W images available for any given cluster (see Table
 \ref{tab:exp}).  Then, at the corresponding positions of the stars
 present in the {\it master list}, a fit was forced with
 \texttt{DAOPHOT/ALLFRAME} \citep{stetson94} in each single frame
 obtained through each filter.  For each star, multiple magnitude
 estimates obtained in each filter were homogenized by using
 \texttt{DAOMASTER} and \texttt{DAOMATCH}, and their weighted mean and
 standard deviation were finally adopted as the star magnitude and
 photometric error.  The final catalog of each cluster has been
 finally cross-correlated with those published in the intermediate release
 \citep{piotto+15} and available upon request at the web page
 \texttt{http://groups.dfa.unipd.it/ESPG/treasury.php}. In all cases,
 we found thousand stars in common, that allowed us to:{\it (i)}
 report the final instrumental magnitudes in each filter to the
 VEGAMAG photometric system, and {\it (ii)} convert the instrumental
 coordinates into the absolute astrometric system.  For NGC 2808, NGC
 7078 and NGC 6388 we used the stellar proper motions (PMs) of
 \citet{bellini+14}, while for NGC 6541 the PM measures available in
 the on-line catalogs at the Survey web page were adopted.

\section{The BSS selection}
\label{bss_box}  
\begin{figure}[h!]
\centering \includegraphics[scale=0.8]{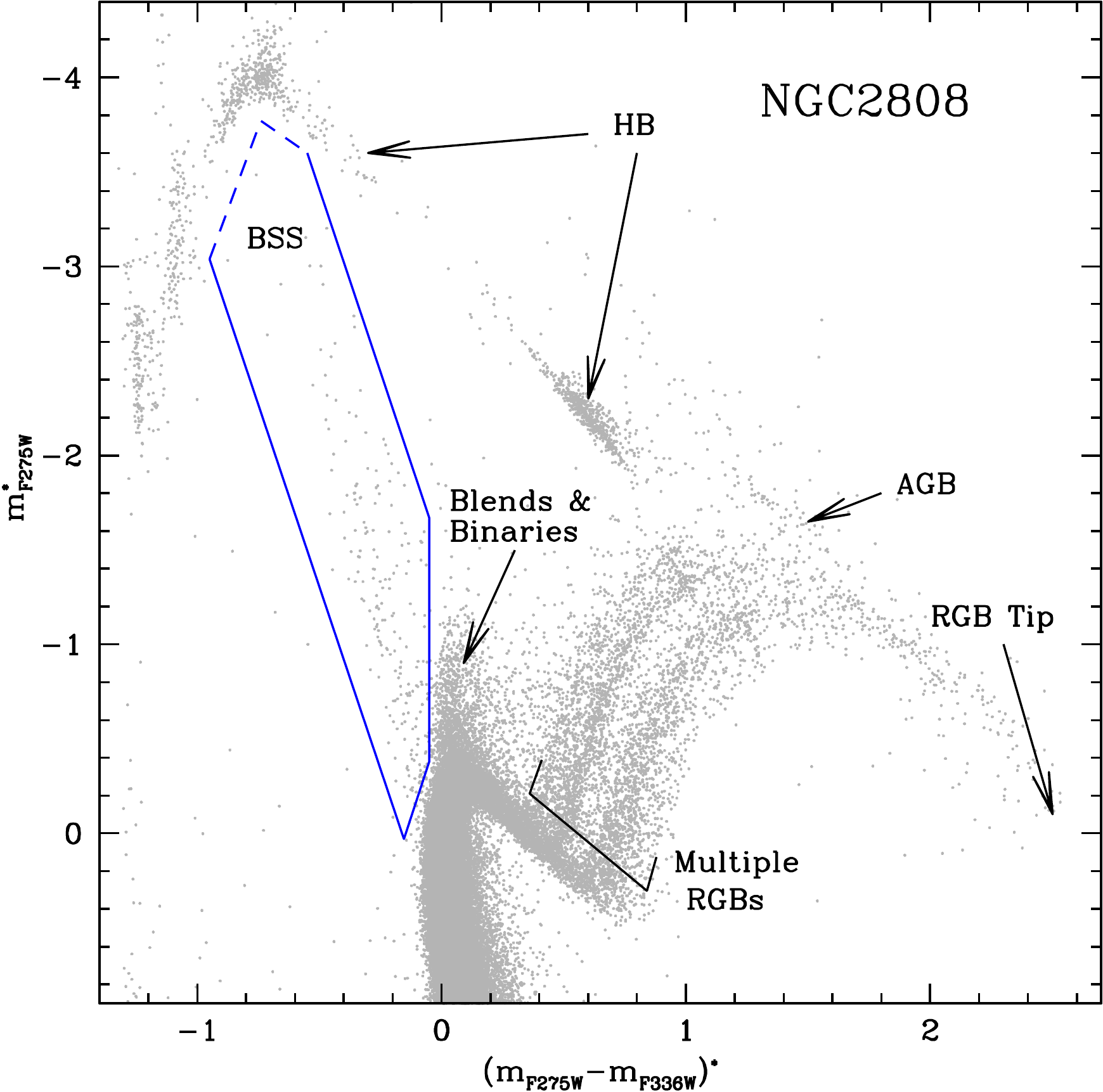}
\caption{``Normalized'' UV CMD of NGC 2808: magnitudes and colors have
  been arbitrarily shifted to locate the MS-TO at $m_{\rm F275W}^*=0$
  and $(m_{\rm F275W}-m_{\rm F336W})^*=0$. The location of the main
  evolutionary sequences is labelled.  Clearly, the brightest objects
  in this diagram are hot horizontal branch (HB) stars and BSSs, while
  cool giants, as red giant branch (RGB) and asymptotic giant branch
  (AGB) stars, are significantly less luminous.  The box adopted
    for the BSS population selection is drawn in blue. The brightest
    boundary of the box is plotted as dashed line, since it can vary
    for cluster-to-cluster depending on the HB morphology (see
    Sect. \ref{bss_box}). The locus expected to be populated by
  photometric blends is also shown.}
\label{cmd2808}
\end{figure}

To define selection criteria for the BSS population that can be
consistently adopted in all clusters (see, e.g., \citealt{leigh07,leigh11}) 
we followed a fully empirical approach. We defined a sort of  ``normalized'' CMD, 
with the magnitudes and the colors arbitrarily shifted to locate the 
MS-TO at approximately
$m_{\rm F275W}^*=0$ and $(m_{\rm F275W} -m_{\rm F336W})^*=0$. 
This allows us to locate the BSS sequence in the same portion 
of the diagram independently of the cluster distance and metallicity. 
Figure \ref{cmd2808} shows the ``normalized'' UV CMD of NGC 2808 (adopted as template). 
For the sake of reference, the position of the main evolutionary sequences 
in the diagram is also labeled. As discussed above, the hot stars belonging 
to the blue tail of the horizontal branch (HB) dominate the UV emission, 
while red and asymptotic giant branch (RGB and AGB) stars are significantly 
fainter. As apparent, this diagram allows an efficient and reliable 
selection of BSS samples. Also the region where 
optical blends (and possibly binaries) are expected to be located is easily 
recognizable in the selected CMD (see the well visible plume 
just above the MS-TO), and their contamination to the BSS 
sample is therefore minimized. 

The selection box has been defined in order to be virtually applicable to all the
clusters observed in the HST UV Legacy Survey. In this diagram, BSSs define a clean sequence
populating a $\sim 2$ mag-wide strip spanning approximately 3 magnitudes, diagonally crossing
the diagram from the cluster MS-TO, toward the blue extension of the HB. Hence the
selection box is defined along two parallel lines including the bulk of the BSS
population. The equations of the two lines are:

$$ m_{\rm F275W}^*= 3.86 \times (m_{\rm F275W} -m_{\rm F336W})^* -1.48$$
$$ m_{\rm F275W}^*= 3.86 \times (m_{\rm F275W}-m_{\rm F336W})^* +0.56 $$

We also define a red boundary to separate BSSs from the supra-MS
plume. A vertical line at $(m_{\rm F275W} -m_{\rm F336W})^* = -0.05$
has been found to efficiently exclude the bulk of the supra-MS plume, which can be more
or less populated, depending on the crowding conditions of each system. The faint edge,
needed to separate BSSs from the MS-TO stars, has been conservatively set at more than 
$5\sigma$ from the mean color of the brightest portion of the MS-TO
stars. The equation defining this edge is
$$ m_{\rm F275W}^*= -4 \times (m_{\rm F275W} -m_{\rm F336W})^* -0.58 $$

In principle a bright/blue edge of the box is needed to distinguish very bright
BSSs from stars populating the blue portion of the HB (when present). However
in order to study the frequency, distribution and luminosity extension of the
BSS population at its bright end, we preferred not to set any formal boundary
on this side of the box. In fact, a precise boundary set on the basis of the
CMD of GCs with blue HB would cause an artificial cut-off of the brightest
portion of the BSS sequence in clusters with red HB morphology. Moreover, we
emphasize that only a few objects are expected in this (very bright) portion of
the BSS sequence and different assumptions on the bright/blue boundary of the
selection box therefore make very little difference. Hence, in clusters with
extended HB morphology, the boundary (see the dashed lines in Figure
\ref{cmd2808}) has
been set by considering the mean colors and the distribution of the HB stars, 
while no formal boundary has been adopted
for red HB systems. Once the entire sample of GCs is analyzed, this approach
will allow us to estimate the fraction of very bright BSSs that can potentially
contaminate the blue portion of the HB in clusters with an extended HB. 

\begin{figure}[h!]
\centering \includegraphics[scale=0.8]{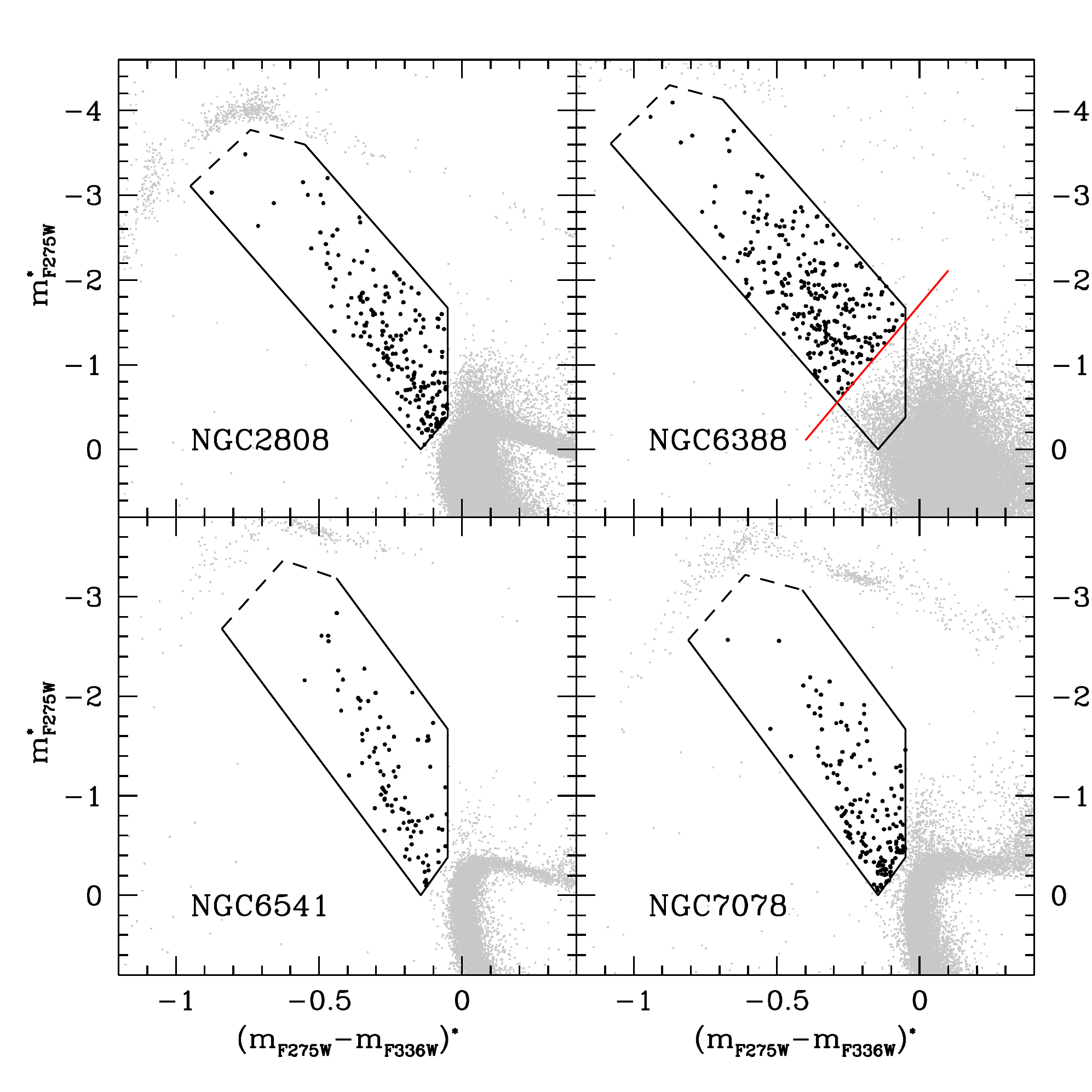}
\caption{Normalized UV CMDs (as in Fig. \ref{cmd2808}) of the four
  program clusters zoomed in the BSS region, with the selected BSS
  populations highlighted with black dots.  The selection box
    defined in Figure \ref{cmd2808} is also plotted in each panel.  In the case of
    NGC 6388 only stars brighter than the red line have been
    considered because of the large photometric errors affecting the
    MS-TO region.}
\label{cmd4}
\end{figure}

The BSS selection box defined according to the above relations is
shown in Figure \ref{cmd2808}. We emphasize that slightly different
assumptions about its boundaries do not alter the main results
presented in this paper. Figure \ref{cmd4} highlights the BSS
populations thus selected in the four program clusters, shown in the
``normalized'' UV CMDs. Note that in the case of NGC 6388, only BSSs
more luminous than $m_{\rm F275W}^*= -4 \times (m_{\rm F275W} -m_{\rm
  F336W})^* -1.71 $ have been considered because of the large
photometric errors affecting the MS-TO region, which prevented a safe
BSS selection at fainter magnitudes. The total number of BSSs selected
in each cluster is listed in Table \ref{tab:bss}.  The final BSS
catalogs will be published in a forthcoming paper (Ferraro et
al. 2017) for all the clusters observed in the HST UV Legacy Survey.

Depending on the time coverage of the observations secured in each cluster, 
in the forthcoming paper we also plan to search for variable BSSs and, 
in the most favorable cases, also publish their light curve. 
This is indeed quite relevant for any BSS study, since some of these stars 
can be eclipsing binaries (in particular, W UMA-type stars; e.g., \citealp{mateo93, rucinski97}) 
or pulsating SX Phoenicis (e.g., \citealp{pych01, fiorentino+14}).

\section{Comparison to previous BSS selections and online catalogs}
\label{compa}
In the recent years, several works aimed at characterizing the
properties of BSSs in GGCs have been published (see, e.g.,
\citealp{leigh07, knigge09, leigh13}). Most of them are based on
online catalogs obtained from the ACS GC Survey
\citep{sarajedini+07}. Unfortunately, those catalogs turn out to be
(sometimes severely) incomplete in the inner regions of high-density
clusters. This is because the ACS GC Survey was designed to secure a
series of dithered and deep images in the optical band (F606W and
F814W filters), with the aim of properly sampling the entire extension
of the MS.  Hence, the bright giant stars are heavily saturated in
those images and their combined blooming effect strongly prevent the
proper measuring of fainter stars in relevant portions of the central
regions of high-density GCs.  In some cases the short exposure (only
one per filter) concurrently acquired in each HST visit alleviates,
but does not completely solve, the problem, since the gap between the
ACS chips remains not sampled at all (and it crosses the central
regions of most of the clusters).  The combination of these effects
produces large incompleteness in the samples of stars measurable above
the MS-TO (including most of the BSSs), and this is well visible even
without specific completeness estimates. In fact, while faint stars
are detected in any cluster region thanks to the dithering pattern
adopted during the observations, the ACS inter-chip gap clearly
appears in the cluster map built by using only stars brighter than the
MS-TO in the ACS GC Survey catalogs of any high-density cluster.
Figure \ref{6388saraj} dramatically demonstrates the problem for the
case of NGC 6388.
Of course any other catalog that has been constructed starting from
the ({\it optical-driven}) ACS GC Survey samples suffers from the same
bias.   Here we discuss a few recent cases.

\subsection{Comparing BSS selection boxes}
\label{leigh}
Based on the ACS GC Survey catalogs, \citet{leigh11} presented a
thoughtful definition of the BSS selection box in the optical
CMD. Unfortunately no direct cluster-to-cluster comparison can be
performed with that work, since no one of the four clusters discussed
here is in common with the sample of \citet{leigh11}. However the
detailed description of how they defined the BSS selection box allows
us to compare our (UV-selected) BSS samples with what would be
obtained by using their criteria.

\begin{figure}[h!]
\centering \includegraphics[scale=0.6]{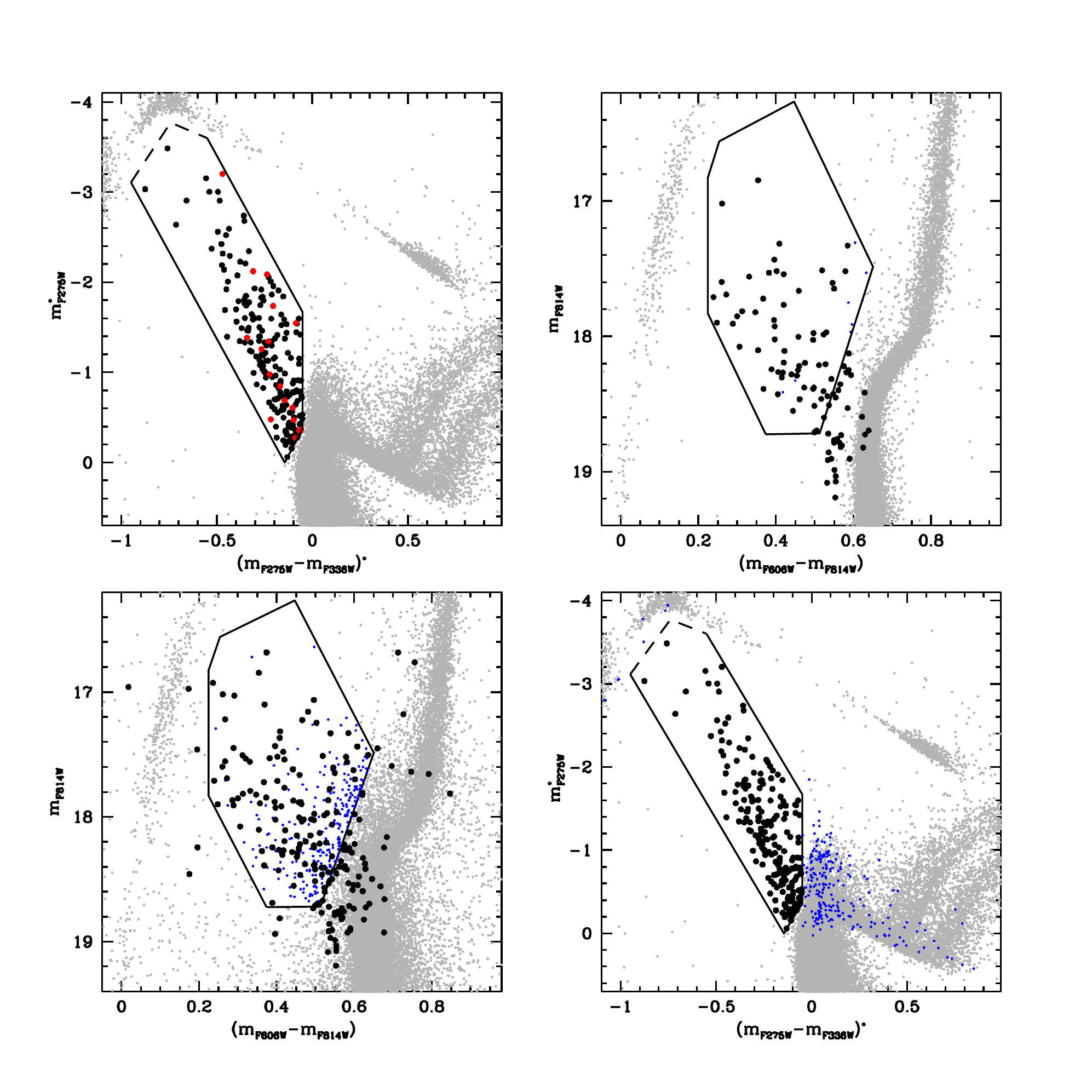}
\caption{ Comparison between UV-based and optical-based BSS
    selections in the case of NGC 2808 (see also Table
    \ref{tab:cfrleigh}).  {\it Upper left panel:} The 215 BSSs selected
    in the UV CMD are shown as large dots.  The 18 sources for which
    no counterpart has been found in the optical catalog are
    highlighted in red.  {\it Bottom left panel:} Optical CMD obtained
    from the ACS GC Survey catalog (\citealt{sarajedini+07}), with the
    BSS selection box built by following the prescriptions of
    \citet{leigh11} drawn. The 197 UV-selected BSSs having an optical
    counterpart are plotted as large black dots.  As it can be seen, a
    significant number of them (84) lie outside the
    BSS optical selection box, while a large number
    (235) of ``intruders'' (small blue dots) is found
    within the BSS optical box.  {\it Bottom-right panel:} UV CMD and
    UV-selected BSSs (as in the upper-left panel), with the position
    of the ``intruders'' marked with small blue dots (as in the
    bottom-left panel).  As it can be seen, in the UV CMD these
    objects are MS/SGB/RGB stars and blends.  {\it Upper-right panel:}
    Optical CMD (same as in the bottom-left panel) obtained by
    considering only stars with good photometry (see
    \citealt{milone+12}). The 95 UV-selected BSSs with good optical
    photometry are shown as large black dots. The optical BSS
    selection box (following \citealt{leigh11}) is also drawn: it
    includes 32 UV-selected BSSs.}
\label{cfrleigh}
\end{figure}

\begin{figure}[h!]
\centering \includegraphics[scale=0.6]{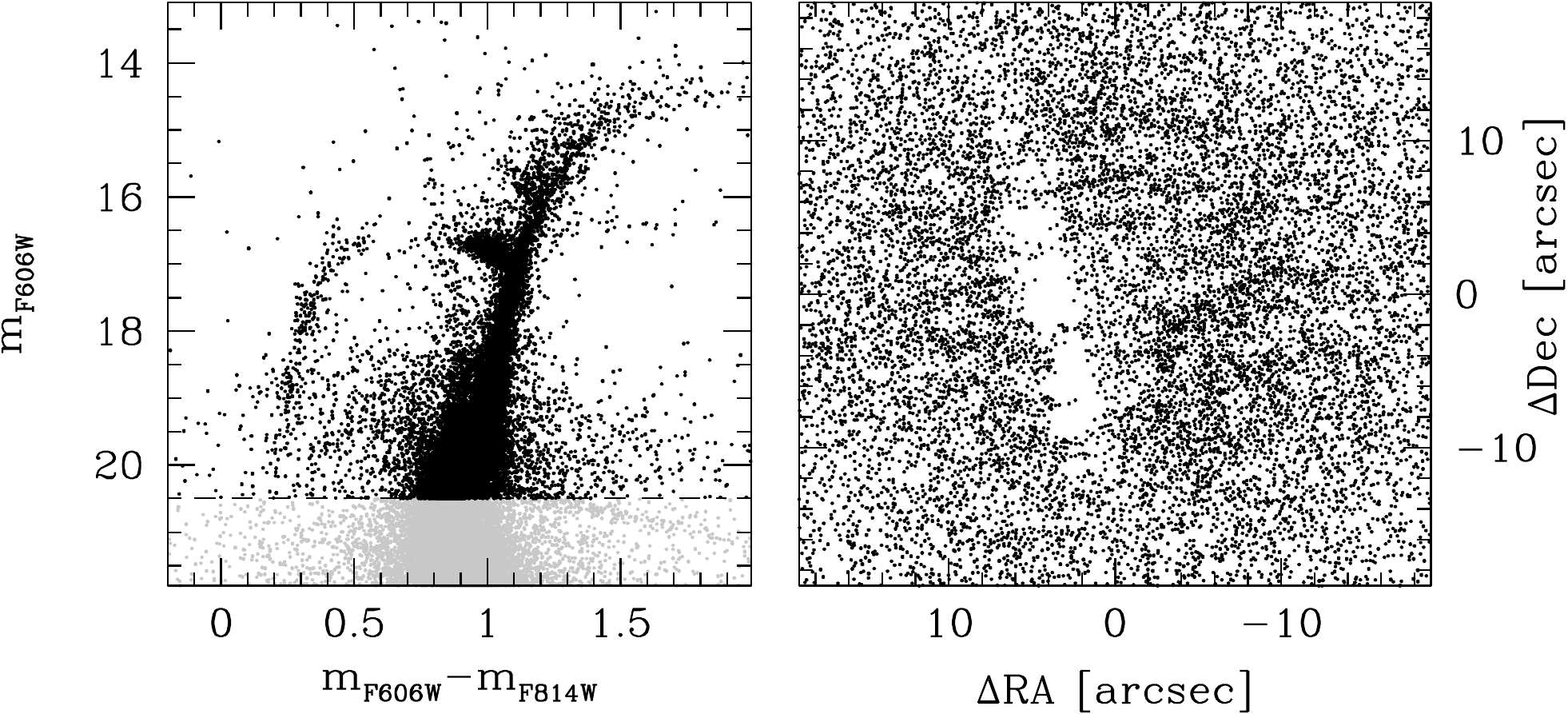}
\caption{{\it Left panel:} optical CMD of NGC 6388 obtained from the
  online ACS GC Survey catalog.  {\it Right panel:} map of the central
  $40\arcsec\times40\arcsec$ of NGC 6388 obtained from all the stars
  brighter than $m_{\rm F606W}=20.5$ in the online ACS GC Survey
  catalog (see black dots above the dashed line in the left
  panel). The empty region is due to the ACS inter-chip gap.}
\label{6388saraj}
\end{figure}

Figure \ref{cfrleigh} and Table \ref{tab:cfrleigh} illustrate the case
of NGC 2808.  In the upper-left panel we show the BSS selection in the
UV diagram, made by using the box discussed in Sect. \ref{bss_box}. A
total number of 215 BSSs is found. Of these, 18 are completely missed
in the optical catalog (see the red dots in the upper-left 
panel of Figure \ref{cfrleigh}), i.e. no
optical counterpart to these sources can be found in the optical data
set. The bottom-left panel of Figure \ref{cfrleigh} shows the optical
CMD of NGC 2808 from the \citet{sarajedini+07} catalog, with the
UV-selected BSSs highlighted as black circles and the \citet{leigh11}
selection box drawn. As apparent, this box includes the bulk of the
UV-selected BSSs  (113 out of 215), indicating that
our selection criteria agree with those of \citet{leigh11}.  
However the comparison also highlights that using the optical CMD (instead of
the UV diagram) can have two opposite consequences: (1) a loss of
genuine BSSs, and (2) the inclusion of several ``intruders'' (i.e.,
stars falling within the optical BSS selection box, that, however, are
blends or MS/SGB/RGB stars in the UV CMD; they are plotted as small blue 
dots in the figure).  
In fact, several (84) UV-selected BSSs lie outside the optical
selection box.  Most of them are located close to the MS-TO, in the
region where the BSS sequence merges into the MS. This shows that the
UV diagram allows a safe exploration of a fainter portion of the BSS
luminosity function, while a proper selection of BSSs in the optical
CMD is limited to brighter objects. A few other stars lie outside the
optical box probably because their optical photometry is perturbed by
the presence of bright giants. In addition, a large number (242) of sources 
that are not UV-selected BSSs is
found within the optical box (blue dots). The location of these
``intruders'' in the UV diagram is shown in the bottom-right panel of
the figure: clearly these are MS/SGB/RGB stars and blends, which
erroneously fall within the BSS selection box in the optical CMD
because of their poor photometry.  The optical CMD obtained by using
only stars with good photometry (see \citealt{milone+12}) is plotted
in the upper-right panel of Figure \ref{cfrleigh} and it shows that,
in fact, essentially all the ``intruders'' (but 7)
have disappeared. This high quality optical CMD confirms that most of
the UV-selected BSSs fall within the \citet{leigh11} box and the
majority of those found beyond the box are located close to the
MS-TO. However, only 62+7 stars would have been
classified as BSSs based on this diagram, while a total of 215 BSSs is
found in the UV. By considering the numbers listed in Table
\ref{tab:cfrleigh} we can conclude that, in the specific case of NGC
2808, an analysis carried out in the optical band would have safely
sampled (by considering only stars with good optical photometry) about
35\% of the UV-selected population
(69 vs 215), while $\sim 15\%$
of them would have been lost because these are too faint to be
properly separated from the MS-TO in the optical. On the other hand,
the entire optical sample (including objects of poor photometry) would
be completely dominated by ``intruders'' (235),
thus making the BSS selection meaningless.

This study clearly highlights the benefits of using UV-driven (instead
of optical-driven) catalogs for the proper selection of BSS samples,
independently of the precise boundaries adopted for the selection
box. It demonstrates that the effect of crowding in the central
regions of high density clusters can (1) prevent the identification of
BSSs that are too close to bright giants (red dots in the upper left
panel), (2) make inaccurate the photometry of genuine BSSs, thus
``moving them out'' of a reasonably drawn selection box (bottom-left
and upper-right panels of Figure \ref{cfrleigh}) and (3) ``move
  within'' the BSS selection box a large number of MS/SGB/RGB stars
  and blends (bottom-right panel of Figure \ref{cfrleigh}).

\subsection{Comparison to online catalogs}
\label{sp16_soto}
\begin{figure}[h!]
\centering \includegraphics[scale=0.6]{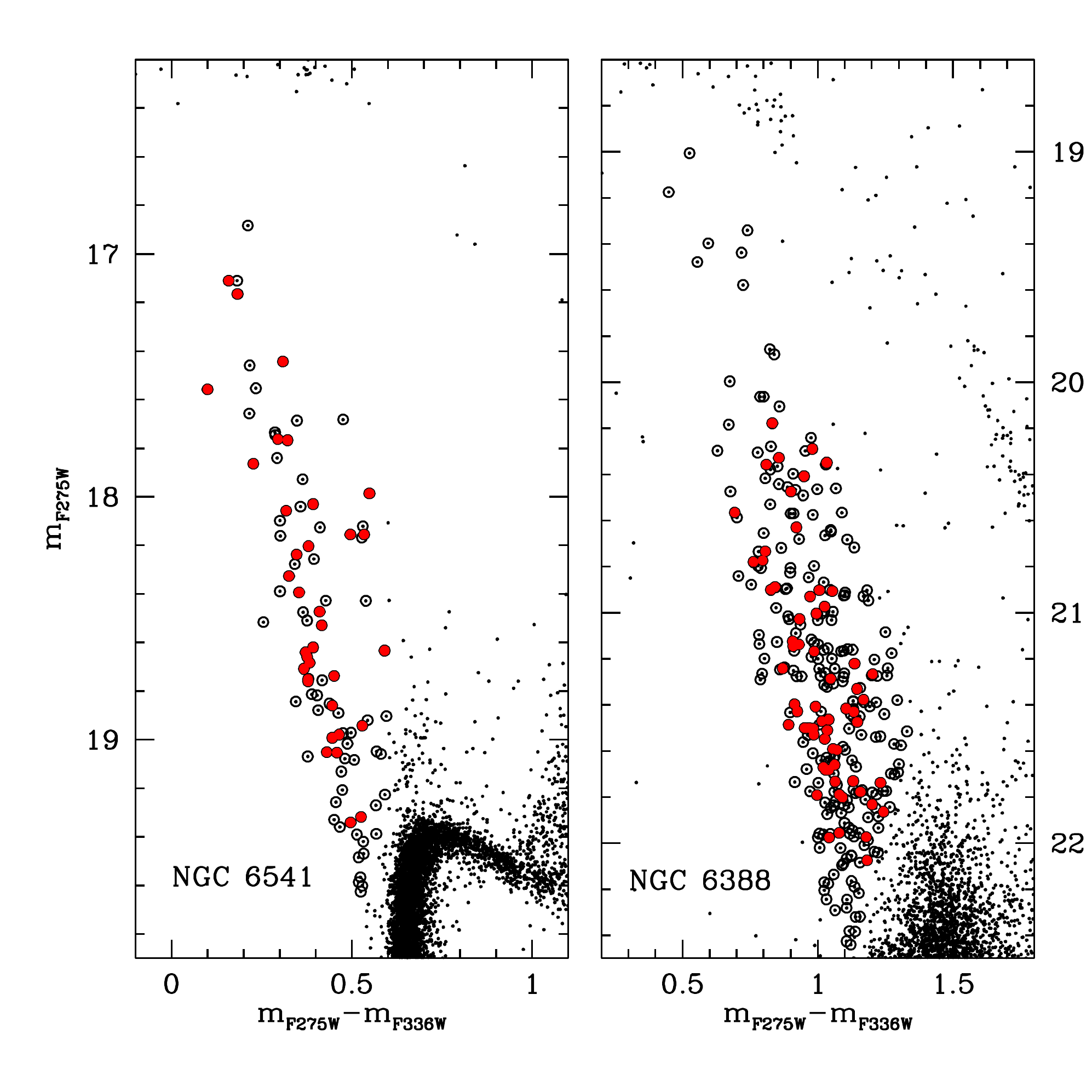}
\caption{UV CMDs of NGC 6541 (left) and NGC 6388 (right), zoomed in
  the BSS region. The large empty circles mark the BSS population
  detected with the {\it UV-guided} approach in this work. The filled
  red circles mark the BSSs also present in the {\it optical-driven}
  and PM-cleaned catalogs of SP16. Clearly, a large fraction of BSSs
  is missed in the latter.}
\label{cfr_simun}
\end{figure}

\citet[][hereafter SP16]{simunovic16} present PM-cleaned catalogs of
BSSs in 35 GGCs. PMs have been measured from the comparison between
the stellar centroids listed in the ACS GC Survey catalogs and those
measured in the F336W exposures of the UV Legacy Survey of
GGCs. Hence, by construction, these catalogs include only stars that
are present in the ({\it optical-driven}) ACS GC Survey sample, and
therefore suffer from the bias described above. In addition, only
stars with high-quality flags have been used, thus
{\it further reducing the number of stars listed in the online
  PM-cleaned BSS catalogs}.  In order to perform a quantitative
comparison, we discuss the specific case of the two clusters in common
with SP16, namely NGC 6541 and NGC 6388. SP16 list 41 BSSs in their
catalog of NGC 6541, also including evolved BSSs (EBSSs), i.e.,
objects that already evolved off the MS phase.  Our {\it UV-driven}
selection identifies 94 BSSs in NGC 6541, and no objects {\it have
  been excluded on the basis of the PM measurements.}  Our sample
includes 35 BSSs in common with SP16, while their remaining 6 stars
correspond, in the UV diagram, to a bright object too close to the HB to
be safely classified as a BSS, and 5 EBSSs.  Thus, a total of 59 BSSs
(representing $63\%$ of the entire sample) are missed in the SP16
catalog (see Figure \ref{cfr_simun}, left panel), possibly because
{\it these objects} did not match the quality-flag criteria adopted by
SP16.
In NGC 6388, our {\it UV-driven} selection identifies 288 candidate
BSSs. Instead, only 74 BSSs and EBSSs are listed in the SP16 online
catalog. Out of these, 64 BSSs are in common with our sample (see
Figure \ref{cfr_simun}, right panel), while the remaining 10 are
located, in the UV diagram, in the region of stellar blends or
EBSSs. Hence, the optical-driven, PM-cleaned and quality fit-selected
sample of BSSs discussed in SP16 includes only a small fraction
($22\%$) of the potential global population of NGC 6388.  About 30
UV-selected BSSs are lacking in the optical sample just because they
fall in the ACS gap (see Figure \ref{6388saraj}).  On the other hand,
the Galactic field contamination is known to be significant in this
cluster and the PM measurements of \citet{bellini+14} suggest that
this is particularly relevant for the faintest portion of the BSS
sequence (see also Figure 6 in SP16).  Unfortunately, the
high-precision PMs currently available for NGC 6388 \citep{bellini+14}
do not include the very center of the cluster, since they are based on
the ACS GC Survey master frame and therefore are also affected by the
presence of the ACS gap (a more sophisticated approach able to include
also the innermost regions of the cluster is planned as part of the
HST UV Legacy Survey of GGCs). Hence, deeper analyses of the BSS
population in NGC 6388 are unfeasible at the moment.  These examples
clearly demonstrate that, while the PM-cleaned BSS catalogs presented
by SP16 allow a secure screening of cluster members useful for
spectroscopic follow-ups, they include only a small fraction of the
cluster BSS population, and thus cannot be used to derive global
properties and perform quantitative studies aimed at exploring, for
example, the BSS radial distribution, central segregation, and the
number ratio of BSS with respect to the cluster ``normal'' stars.

\begin{figure}[h!]
\centering \includegraphics[scale=0.7]{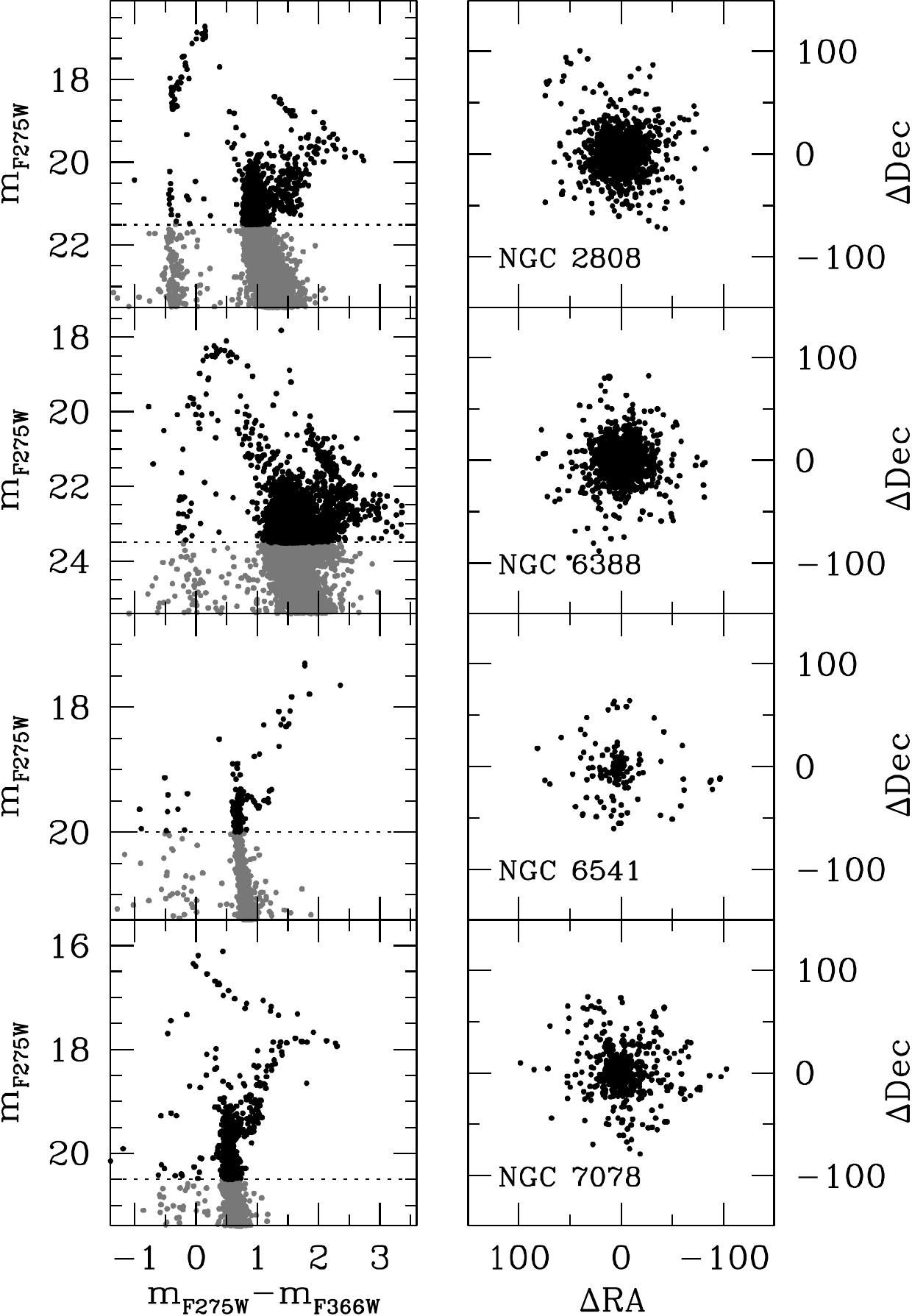}
\caption{{\it Left:} UV CMDs of the stars missed in the optical-driven
  catalogs of \citet{soto16} for the four program clusters.  {\it
    Right:} spatial distribution (in arcseconds and with respect to
  the cluster center) of the missed stars brighter than the limits
  marked by the dashed lines in the left panels.  The number of the
  missed stars is labelled in each panel.}
\label{cfr_soto}
\end{figure}

\begin{figure}[h!]
\centering \includegraphics[scale=0.7]{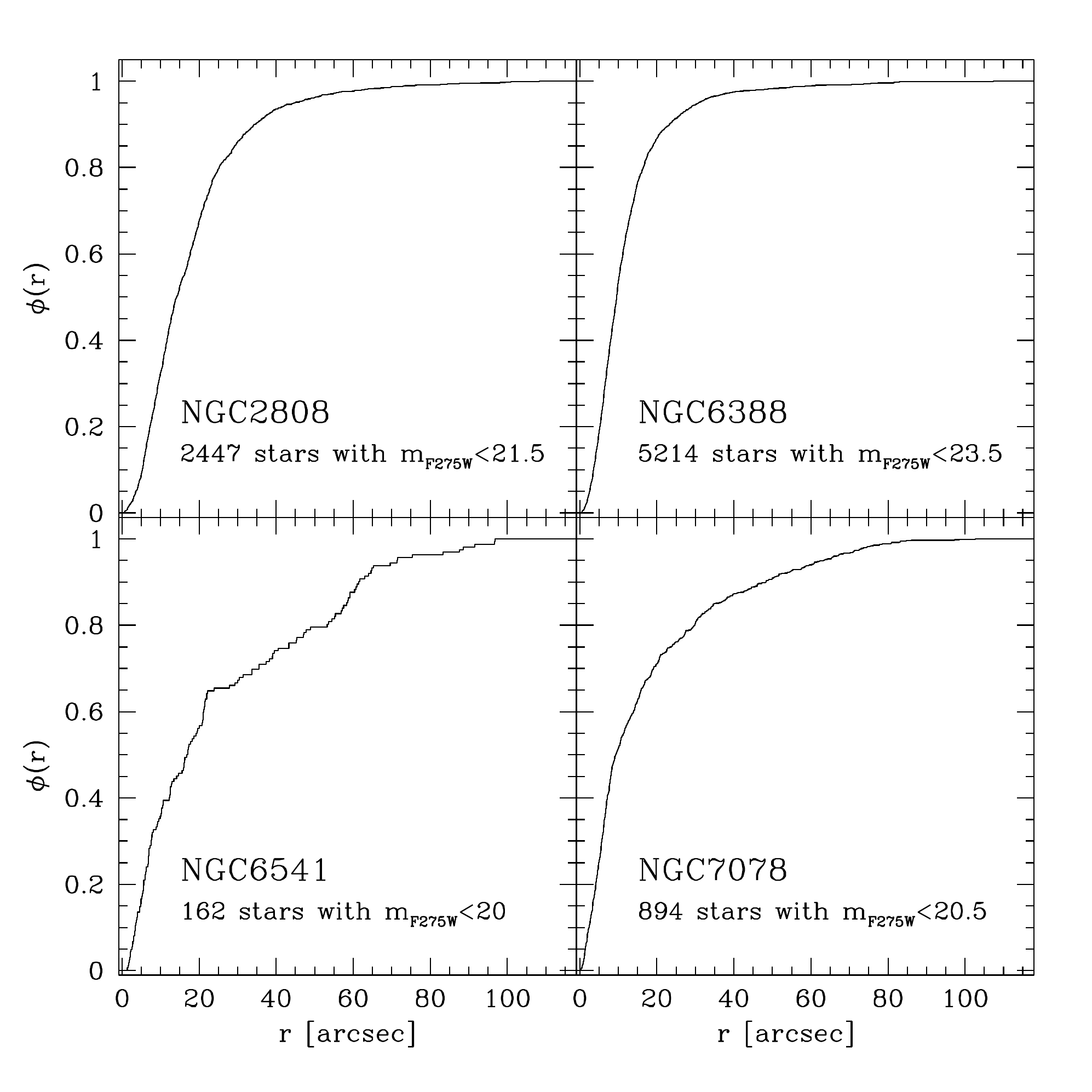}
\caption{Cumulative radial distribution of the bright stars (brighter
  than the dashed lines in the left panels of Figure \ref{cfr_soto})
  missed in the optical-driven catalogs \citep{soto16} of the four
  program clusters. The central segregation of the missed stars is
  evident: in all the cases a fraction larger than $\sim 70\%$  of missed stars
  is located within the innermost $10\arcsec$ from the cluster center.}
\label{radistsoto}
\end{figure}

\citet{soto16} present a preliminary public release of photometric
catalogs for the 57 GGCs observed in the UV Legacy Survey of GGCs.
These catalogs have been built by assigning the magnitudes measured in
the F275W, F336W and F438W filters to all the stars listed in the
({\it optical-driven}) ACS GC Survey samples. Hence, by construction,
also these catalogs suffer from the problems discussed above and are
not suitable for quantitative studies of the evolved stellar
populations (brighter than the MS-TO).  The detailed comparison
between the Soto online catalogs and the ``UV-guided'' samples
discussed in this paper offers the opportunity to quantitatively study
the distribution (both in space and in the CMD) of the stars missed in
the {\it optical-driven} approach.  For the four clusters under study,
Figure \ref{cfr_soto} shows the UV CMD and the spatial distribution of
all the stars detected in the {\it UV-guided} approach that are missed
in the Soto online catalogs.  As apparent, they are distributed along
all the evolved sequences in the CMD (HB, RGB, BSS and white dwarf
sequences) and a significant number of stars is also missed in the
MS-TO and upper MS regions.  The number of missed stars depends on the
cluster structure: by considering only the brightest portion of the
CMD (approximately brighter than the cluster MS-TO), it varies from
near 200 in NGC 6541, up to more than five thousand in NGC 6388.  The
right panels show the spatial distribution of the missed stars,
revealing that they are mainly concentrated in the innermost regions,
as it is expected in the case of incompletness due to an effect of
stellar crowding.  To further illustrate the problem, in Figure
\ref{radistsoto} we plot the cumulative radial distribution of the
stars lost in the optical-search for the four program clusters.  As it
can be seen, the vast majority ($\gtrsim$ 70\%) of the missed stars is located
within the innermost $20\arcsec$-$30\arcsec$ from the cluster center,
thus demonstrating the benefits of using UV-driven over optical-driven
  catalogs to study the radial distributions and
population ratios of evolved stars.
This analysis shows also that the online {\it optical-driven} catalogs should be used with
caution when quantitative studies of any evolved stellar population is
attempted.

\section{Measuring the level of radial segregation of BSSs}
\label{apiu}
As discussed above, BSSs are a population of heavy stars, with an
average mass $\langle m_{\rm BSS}\rangle = 1.2 M_\odot$
\citep{shara97, gilliland98, ferraro+06b, fiorentino+14, baldwin+16},
orbiting a ``sea" of lighter objects (the average stellar mass in old
GCs is $\langle m \rangle\sim 0.3 M_\odot$). For this reason they are
subject to the action of dynamical friction that make them segregate
toward the cluster center, with an efficiency that decreases for
increasing radial distance (e.g., \citealt{BT87, mapelli04,
  mapelli06}).  \citet{ferraro+12} demonstrated that this process is responsible for
the observed shapes of the BSS normalized radial
distribution\footnote{The ``normalized BSS distribution''
  \citep{ferraro+93} is defined as the ratio between the fraction of
  BSSs sampled in any adopted radial bin and the fraction of cluster
  light sampled in the same bin.}  (hereafter BSS-nRD), that in most
of the analyzed clusters is bimodal, with a central peak, a minimum at
a distance $r_{\rm min}$ from the center, and an external rising
branch.  The position of the minimum, expressed in units of the
cluster core radius, has been used by \citet{ferraro+12} as a clock-hand to measure
the dynamical ages of GCs. In fact, being all clusters coeval (same
chronological age, $t\sim 12-13$ Gyr), dynamical friction acted on
their BSS population during the same period of time. However,
different shapes of the BSS-nRD are currently observed, with $r_{\rm
  min}$ spanning the entire radial range, from the very center out to
the outskirts, depending on the system. Hence, dynamical friction has
been more or less efficient in segregating BSSs toward the center in
different GCs, and $r_{\rm min}$ is an indicator of this action: small
(large) values of $r_{\rm min}$ are found in dynamically young (old)
clusters, while intermediate values are measured in sub-families of
intermediate dynamical-age systems.

Under these assumptions, the outward propagation of the BSS-nRD
minimum is expected to be linked to a growing level of BSS segregation
in the central regions of the cluster. In order to explore this
aspect, \citet{alessandrini+16} and \citet{lanzoni+16} proposed to
measure the level of BSS segregation by means of the parameter $A^+$,
defined as the area enclosed between the cumulative radial
distribution of BSSs, $\phi_{\rm BSS}(x)$, and that of a reference
(lighter) population, $\phi_{\rm REF}(x)$:

\begin{equation}
  A^+(x) = \int_{x_{\rm min}}^x \phi_{\rm BSS}(x') -\phi_{\rm REF}(x') ~dx',
\label{eq_A+} 
\end{equation}

where $x=\log(r/r_h)$ is the logarithm of the distance from the
cluster center normalized to the cluster half-mass radius $r_h$.  Note
that the distances from the cluster center are expressed in
logarithmic units in order to maximize the sensitivity of $A^+$ to the
BSS segregation (which is prominent in the central regions of each
cluster). Indeed, by means of direct $N-$body simulations, 
\citet{alessandrini+16}
demonstrated that the value of $A^+$ systematically increases with
time, as expected for a sensitive indicator of the BSS sedimentation
process. 
\citet{lanzoni+16} discussed the operational procedure to measure
$A^+$ in observational samples and emphasized that in order to perform
a meaningful cluster-to-cluster comparison, the parameter must be
measured over equivalent radial portions in every system. For this
reason, the value of $A^+$ was measured within one half-mass radius
($r_h$) from the center (i.e., out to $x=0$) in a sample of 25 GGCs,
and a tight link between the parameter (hereafter $A^+_{rh}$) and the
position of the BSS-nRD minimum was found \citep{lanzoni+16},
providing an empirical link between the two indicators and confirming
that both are describing the same physical process.  In addition
\citet{lanzoni+16} found a nice correlation between $A^+_{rh}$ and the
central relaxation time of the cluster ($t_{rc}$), in the sense that
the new indicator systematically decreases for increasing values of
the relaxation time. Thus, the parameter $A^+_{rh}$ turns out to be an
efficient indicator of the level of dynamical evolution experienced by
star clusters since their formation.

NGC 6388 is one of the clusters discussed in \citet{lanzoni+16}, who
adopted the BSS sample obtained from HST-WFPC2 observations by
\citet{dalessandro+08}, which was complete but limited only to the
brightest portion of the sequence (corresponding to $m_{\rm
  F275W}<21.5$).\footnote{Note that the sample of BSSs used in
  \citet{lanzoni+16} to measure the parameter $A^+_{rh}$ was the same
  adopted by \citet{ferraro+12} to locate the position of the minimum
  of the BSS-nRD. This required to cover the entire cluster radial
  extension by complementing HST data with wide-field photometry
  obtained from the ground. In turn, the necessity of considering
  samples of BSSs homogeneous in magnitude over the whole cluster
  forced us to use only the brightest portion of the BSS sequence even
  in the regions observed with the HST, where much fainter magnitudes
  were reached.\label{foot:bssbri}} Indeed, the value of $A^+_{rh}$
that we measure here by considering only such a sub-sample turns out
to be $A^+_{rh}=0.19$, in perfect agreement with the value quoted in
\citet{lanzoni+16}. On the other hand, the severe Galactic field
contamination affecting the faintest portion of the BSS sequence (see
also SP16) and the lack of accurate PMs especially in the innermost
cluster regions (see Sect. \ref{sp16_soto}), suggest that the
investigation of the entire BSS sequence of NGC 6388 is unreliable at
the moment.  Hence for this cluster here we adopt the value
$A^+_{rh}=0.19$.

For the other three clusters in our samples, we followed the
prescriptions by \citet{lanzoni+16} to measure the value of
$A^+_{rh}$. As a first step, the cluster center needs to be
determined. To this end, we adopted the iterative procedure proposed
by \citet[][see also \citealt{lanzoni+10} and
  \citealt{miocchi+13}]{montegriffo+95}, and we determined the center
of gravity ($C_{\rm grav}$) by averaging the right ascension
($\alpha$) and declination ($\delta$) of all the stars detected within
a circle of (first-guess) radius $r$, brighter than a given magnitude
limit. Depending on the cluster characteristics, in each GC we
selected the optimal range of stellar magnitudes to guarantee high
enough statistics and to avoid spurious effects due to photometric
incompleteness.  In the specific cases presented here, we used all the
stars down to the MS-TO region (approximately with $m_{F275W}^*<0.8$).
Table \ref{centers} lists the adopted values of $C_{\rm grav}$ and
reports the differences with respect to those quoted in
\citet{goldsbury+10}.  These are always smaller than $0.35\arcsec$,
both in $\alpha$ and in $\delta$, with the exception of NGC 6541, for
which a difference of $\sim 1.8\arcsec$ in $\alpha$ is found. 

Once $C_{\rm grav}$ is located, to determine $A^+_{rh}$ one needs to
compare the cumulative radial distribution of BSSs out to $r_h$, with
that of a reference population assumed to trace normal cluster stars.
In all the selected GCs, the radial region extending from $C_{\rm
  grav}$ out to $r_h$ is entirely covered by the WFC3/UVIS data.  For
a proper comparison with the results of \citet{lanzoni+16}, where only
the bright portion of the BSS sequence has been used (see footnote
\ref{foot:bssbri}), here we computed $A^+_{rh}$ by considering BSSs
with $m_{\rm F275W}^* < -1.5$ (see Fig. \ref{cmd4}).  Moreover, we
adopted both the HB and the RGB populations as reference, and we
assumed the average of the two resulting values as the best estimate
of $A^+_{rh}$, and the standard deviation as its error.

\begin{figure}[h!]
\centering \includegraphics[scale=0.7]{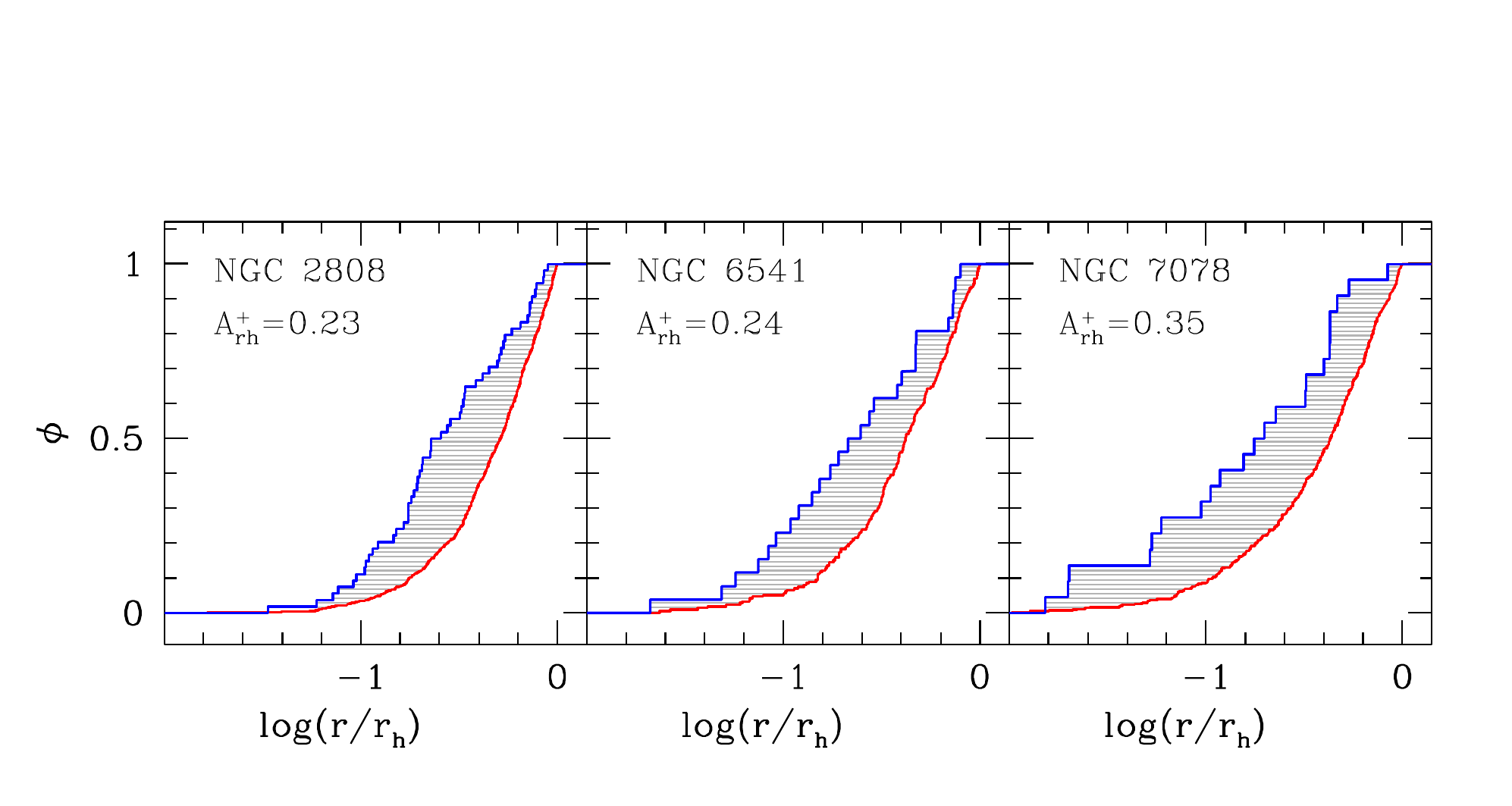}
\caption{Cumulative radial distributions of BSSs (blue line) and
  reference stars (red line) observed within one half-mass radius
  ($r_h$) in NGC 2808, NGC 6541, NGC 7078.  The area between the two
  curves (shaded in grey) corresponds to the labelled value of
  $A^+_{rh}$ (see also Table \ref{tab:bss}).}
\label{cumu}
\end{figure}

\begin{figure}[h!]
\centering \includegraphics[height=15.0cm, width=15.0cm]{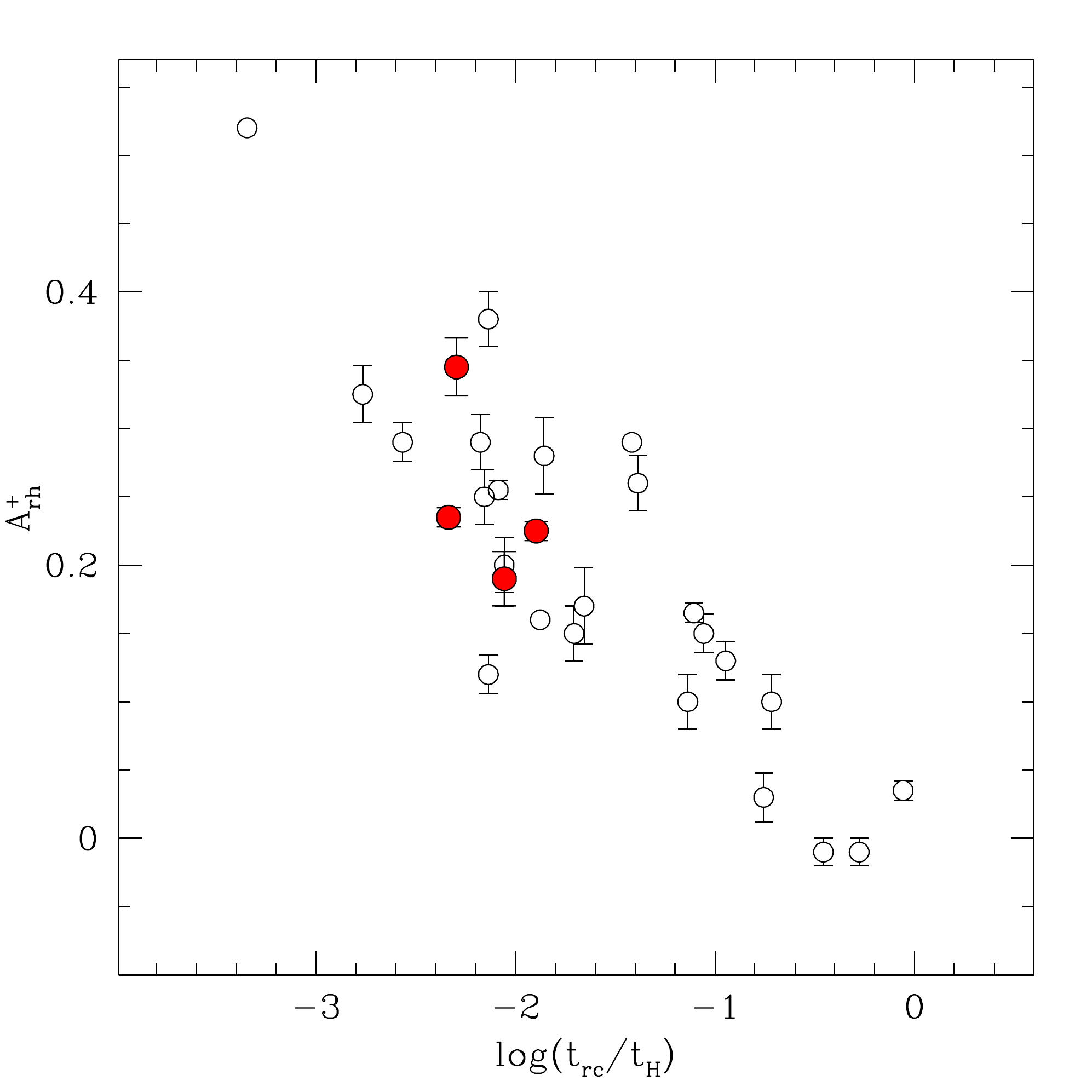}
\caption{Relation between the parameter $A^+_{rh}$ and the cluster
  central relaxation time ($t_{\rm rc}$) normalized to the Hubble time
  ($t_H=13.7$ Gyr). The grey circles are the 25 clusters presented in
  \citet{lanzoni+16}, while the clusters discussed here are plotted as
  red circles.}
\label{apiu_ltrc}
\end{figure}

Figure \ref{cumu} shows the cumulative radial distributions of BSSs
and HB stars for the clusters in our sample.  The adopted values
of $A^+_{rh}$ are labelled in the figure and listed in Table
\ref{tab:bss}.\footnote{The values of $A^+_{rh}$ obtained from the
  catalogs of \citet{soto16} and \citet{simunovic16} for the clusters
  in common are always smaller than those found here, with average and
  maximum differences of 0.04 and 0.09, respectively.}  In the case of
NGC 6541, the value of $A^+_{rh}$ changes from 0.24 to 0.22 if the
center determined by \citet{goldsbury+10} is adopted. More in general,
to explore the effect of an unprecise location of the cluster center,
we have re-determined $A^+_{rh}$ by applying an arbitrary shift of
$0.5\arcsec-1\arcsec$ in all directions to the nominal value of
$C_{\rm grav}$. We found variations of 0.01-0.04, the largest effect
beeing observed (as expected) for the most concentrated system (NGC
7078).  Figure \ref{apiu_ltrc} shows the behaviour of the $A^+_{rh}$
parameter as a function of $t_{rc}$ (expressed in units of Hubble time
$t_H$) for the four clusters presented here (see red circles),
compared to the 25 GGCs discussed in \citet{lanzoni+16}.  As it can be
seen, they nicely fit into the relation, thus providing new support to
the fact that $A^+_{rh}$ is a powerful indicator of the cluster
dynamical evolution and, once properly calibrated, it promises to be
usable as a direct measure of the central relaxation time of these
systems.

\section{Summary}
The results discussed in this paper represent the first application of
the so-called {\it UV-route to the BSS search} for the targets
observed in the UV Legacy Survey of GGCs. As a preparatory work, here
we discussed the {\it UV-driven} BSS catalogs of four massive and
high-density clusters, addressing the following topics: {\it (i)} we
quantitatively demonstrated that the {\it UV-driven} search is the
most efficient route to construct complete samples of BSSs; {\it (ii)}
we showed the benefits of using UV-driven over optical-driven
  catalogs for the quantitative analysis of evolved stellar
populations; {\it (iii)} we provided additional support to the use of
the $A^+_{rh}$ parameter as a powerful indicator of cluster dynamical
evolution.  In a forthcoming paper (Ferraro et al. 2017) we will
extend this analysis to the entire sample of clusters observed in the
UV Legacy Survey, presenting and discussing the final catalogs of BSSs
for all the observed systems and the calibration of the parameter
$A^+_{rh}$ in terms of the parent cluster dynamical age.

\acknowledgements{}

\newpage
\begin{table}[h!]
\begin{center}
\begin{tabular}{lcc}
\hline
Filter &	 F275W & F336W   \\
Name     & $n\times t_{\rm exp}$ & $n\times t_{\rm exp}$      \\
\hline
NGC 2808 & $12 \times 985$ & $6 \times 650$     \\

NGC 6388 & $2 \times 888$ & $4 \times 350$   \\
 		 & $ 2 \times 889$ &  			   \\
		 & $ 2 \times 961$ &      \\
 		 & $ 2 \times 999$ &      \\

NGC 6541 & $2 \times 708$ & $4 \times 300$   \\
		  & $2 \times 758$ & 				  \\
		  
NGC 7078 & $3 \times 615$ & $6 \times 350$   \\
		  & $3 \times 700$ & 		   \\
\hline
\end{tabular}
\end{center}
\caption{Number of exposures ($n$) and exposure times in seconds
  ($t_{\rm exp}$) of the data used to study the BSS population in the
  four program clusters.}
\label{tab:exp}
\end{table}

\newpage
\begin{table}[h!]
\begin{center}
\begin{tabular}{lccccc}
\hline
 &  & &  & \\
Name &  $N_{\rm BSS}$ &  $r_h$ & $\log(t_{\rm rc})$   & $A^+_{rh}$ & $\epsilon$ \\
&  & & & & \\
\hline
NGC 2808    & 215   &   48.0 &  8.24  &   $0.23$ &  0.01 \\
NGC 6388    & 288   &   45.4 &  8.08  &   $0.19$ &  0.02 \\
NGC 6541    &  94   &   63.6 &  7.80  &   $0.25$ &  0.01 \\
NGC 7078    & 167   &   60.0 &  7.84  &   $0.34$ &  0.02 \\
\hline
\end{tabular} 
\end{center}
\caption{Parameters of the program clusters: total number of BSSs
  detected in the WFC3 field of view (column 1); half-mass radius in
  arcseconds (column 2), logarithm of the central relaxation time in
  Gyr (column 3), value of $A^+_{rh}$ and its error (columns 5 and
  6).}
\label{tab:bss}
\end{table}

\newpage
\begin{table}[h!]
\begin{center}
\begin{tabular}{lcccc}
\hline
 &  & &  & \\
Sample  &  $N_{\rm UV-BSS}$ &  $N_{\rm UV-BSS,IN}$  & $N_{\rm UV-BSS,OUT}$ &   Intruders \\
&  & & &  \\
\hline
Good opt. photometry    & 102   & 62 &  32 &  7 \\
Poor opt. photometry    &  95  & 51  & 52   & 235  \\
No  opt. counterpart    &  18  &  - &  -  & - \\
\hline
Total   & 215  & 113  &  84 & 242  \\
\hline
\end{tabular} 
\end{center}
\caption{ Number of UV-selected BSSs with good optical photometry
    (first row, see  \citealt{milone+12}), poor optical photometry (second row), and no optical
    counterpart (third row), found within the UV selection box ($N_{\rm UV-BSS}$, column
    2), within the optical selection box ($N_{\rm UV-BSS,IN}$, column 3), and outside the
    optical selection box ($N_{\rm UV-BSS,OUT}$, column 4). The last column lists the number
    of ``intruders'', i.e., stars that are found within the optical
    selection box but are not selected as BSSs in the UV diagram.  The
    optical selection box has been defined following the prescription
    of \citet{leigh11}. See Sect. \ref{leigh} for a detailed
    discussion. }
\label{tab:cfrleigh}
\end{table}

\newpage
\begin{table}[h!]
\begin{center}
\begin{tabular}{lrrrrrr}
\hline & & & & \\ Name & $\alpha$ & $\delta$ & $\alpha$ (Goldsbury) &
$\delta$(Goldsbury) & $\Delta\alpha$ & $\Delta\delta$ \\
& (h:m:s) & $(\deg:\arcmin:\arcsec)$ & (h:m:s) & $(\deg:\arcmin:\arcsec)$ & (arcsec) & (arcsec) \\
& & & & & & \\
\hline
NGC 2808 & 09:12:03.08 & $-64$:51:48.80 & 09:12:03.10 & $-64$:51:48.6 & $-0.34$ & $-0.20$ \\ 
NGC 6388 & 17:36:17.21 & $-44$:44:07.64 & 17:36:17.23 & $-44$:44:07.8 & $-0.34$ & $ 0.16$ \\
NGC 6541 & 18:08:02.48 & $-43$:42:53.42 & 18:08:02.36 & $-43$:42:53.6 &   1.80  &   0.18  \\
NGC 7078 & 21:29:58.31 &    12:10:01.30 & 21:29:58.33 &    12:10:01.2 & $-0.31$ &   0.09  \\
\hline
\end{tabular} 
\end{center}
\caption{Centers of gravity determined here (columns 2 and 3),
  compared to those published in \citet{goldsbury+10} (columns 4 and
  5). The differences (in arcsec) between the two determinations of
  $\alpha$ and $\delta$ are listed in columns 6 and 7.}
  \label{centers}
\end{table}

\newpage

\newpage

\newpage

\newpage

\newpage

\newpage
 
\newpage

\newpage

\newpage
 
\end{document}